\documentclass[twocolumn]{aastex61}

\usepackage[T1]{fontenc}
\usepackage{bm}

\newcommand{\RN}[1]{\textup{\uppercase\expandafter{\romannumeral#1}}}

\submitjournal{ApJ}

\begin{document}

\title{The Post-Pericenter Evolution of the Galactic Center Source G2}

\correspondingauthor{P. M. Plewa}
\email{pmplewa@mpe.mpg.de, ste@mpe.mpg.de}

\author{P. M. Plewa}
\affil{Max-Planck-Institut f\"ur Extraterrestrische Physik, D-85748 Garching, Germany}

\author{S. Gillessen}
\affil{Max-Planck-Institut f\"ur Extraterrestrische Physik, D-85748 Garching, Germany}

\author{O. Pfuhl}
\affil{Max-Planck-Institut f\"ur Extraterrestrische Physik, D-85748 Garching, Germany}

\author{F. Eisenhauer}
\affil{Max-Planck-Institut f\"ur Extraterrestrische Physik, D-85748 Garching, Germany}

\author{R. Genzel}
\affil{Max-Planck-Institut f\"ur Extraterrestrische Physik, D-85748 Garching, Germany}
\affil{Department of Physics and Astronomy, University of California, Berkeley, CA-94720, USA}

\author{A. Burkert}
\affil{Max-Planck-Institut f\"ur Extraterrestrische Physik, D-85748 Garching, Germany}
\affil{Universit\"ats-Sternwarte M\"unchen, Scheinerstra{\ss}e 1, D-81679 M\"unchen, Germany}

\author{J. Dexter}
\affil{Max-Planck-Institut f\"ur Extraterrestrische Physik, D-85748 Garching, Germany}

\author{M. Habibi}
\affil{Max-Planck-Institut f\"ur Extraterrestrische Physik, D-85748 Garching, Germany}

\author{E. George}
\affil{Max-Planck-Institut f\"ur Extraterrestrische Physik, D-85748 Garching, Germany}

\author{T. Ott}
\affil{Max-Planck-Institut f\"ur Extraterrestrische Physik, D-85748 Garching, Germany}

\author{I. Waisberg}
\affil{Max-Planck-Institut f\"ur Extraterrestrische Physik, D-85748 Garching, Germany}

\author{S. von Fellenberg}
\affil{Max-Planck-Institut f\"ur Extraterrestrische Physik, D-85748 Garching, Germany}

\begin{abstract}
In early 2014 the fast-moving near-infrared source G2 reached its closest approach to the supermassive black hole Sgr~A* in the Galactic Center. We report on the evolution of the ionized gaseous component and the dusty component of G2 immediately after this event, revealed by new observations obtained in 2015 and 2016 with the SINFONI integral field spectrograph and the NACO imager at the ESO~VLT. The spatially resolved dynamics of the Br$\gamma$ line emission can be accounted for by the ballistic motion and tidal shearing of a test-particle cloud that has followed a highly eccentric Keplerian orbit around the black hole for the last 12 years. The non-detection of a drag force or any strong hydrodynamic interaction with the hot gas in the inner accretion zone limits the ambient density to less than a few ${10^3~\mathrm{cm^{-3}}}$ at the distance of closest approach (${1500~R_s}$), assuming G2 is a spherical cloud moving through a stationary and homogeneous atmosphere. The dust continuum emission is unresolved in L'-band, but stays consistent with the location of the Br$\gamma$ emission. The total luminosity of the Br$\gamma$ and L' emission has remained constant to within the measurement uncertainty. The nature and origin of G2 are likely related to that of the precursor source G1, since their orbital evolution is similar, though not identical. Both object are also likely related to a trailing tail structure, which is continuously connected to G2 over a large range in position and radial velocity.
\end{abstract}


\section{Introduction}
\label{sec:1}

High-resolution near-infrared observing techniques have enabled ever more detailed studies of the Galactic Center black hole, which has been identified with the compact radio source Sgr~A*, and its immediate environment \citep[for a review, see][]{2010RvMP...82.3121G}. Almost all of the detectable sources within the central light year are stars. However, a few sources show both continuum emission in L-band and recombination emission in hydrogen and helium lines, but are missing a counterpart source in K-band \citep{2005A&A...439L...9C,2005ApJ...635.1087G,2013A&A...551A..18E,2014AAS...22323805S}. These sources do not resemble the surrounding stars, but appear to be concentrations of gas and dust that are being ionized by the radiation of the nearby massive young stars. One of these objects, called G2, is of special interest due to its proximity to the central supermassive black hole. G2 was discovered in 2011 \citep{2012Natur.481...51G} and was found to be on a highly elliptical orbit, which would bring it to a point of closest approach to Sgr~A* in early 2014 \citep{2013ApJ...773L..13P}. We have been able to follow the orbital evolution of G2 closely by means of imaging spectroscopy, conducting regular monitoring observations and using archive data from as early as 2004 \citep{2013ApJ...763...78G,2013ApJ...774...44G,2015ApJ...798..111P}. Most notably, in position-velocity space G2 became ever more elongated along its orbit as it moved towards Sgr~A* and its radial velocity started to change from being strongly red-shifted to blue-shifted in 2013. The next year's observations captured G2 in the very middle of its extended pericenter passage.
  
During this time G2 has brought a substantial amount of mass to a distance of just a few thousand times the Schwarzschild radius of Sgr~A* ($R_S$). The estimated gas mass of G2 is ${3~M_\Earth}$ \citep{2012Natur.481...51G,2014ApJ...783...31S}, which is comparable to the mass of gas in the hot accretion flow at this distance \citep{2003ApJ...598..301Y,2006ApJ...640..319X}. G2 thus represents a unique probe of the inner accretion zone around the black hole. A possible sign of an interaction between G2 and the ambient gas could have been the detection of X-ray or radio emission originating from a shock front \citep{2012Natur.481...51G,2012ApJ...757L..20N,2013MNRAS.432..478S}. However, the level of observable emission depends sensitively on the poorly known physical parameters of G2 and the accretion flow, for example the densities and volume filling factors. It is thus not surprising that no such emission has been detected \citep[e.g.][]{2015ApJ...802...69B}. Any direct accretion of material onto Sgr~A* would set in only years after the time of pericenter, as earlier simulations have shown \citep[e.g.][]{2012ApJ...755..155S}.

The astrophysical nature of G2 is still debated. Many models and formation scenarios have been proposed, including a clumpy stream of gas that formed from a collision of stellar winds \citep{2006MNRAS.366..358C,2012ApJ...750...58B,2012Natur.481...51G,2012ApJ...755..155S,2015ApJ...811..155S,2016MNRAS.455.4388C} or from the partial tidal disruption of a star \citep{2014ApJ...786L..12G}, different kinds of evaporating circumstellar disks \citep{2012NatCo...3E1049M,2012ApJ...756...86M} or (proto-)planets \citep{2015ApJ...806..197M}, different kinds of enshrouded stars \citep{2012A&A...546L...2M,2013ApJ...768..108S,2013ApJ...776...13B,2014ApJ...789L..33D,2015ApJ...800..125V,2016ApJ...819L..28B} and a binary star merger product \citep{2014ApJ...796L...8W}. The data so far only provides direct evidence for an extended cloud of gas and associated dust (Sec.~\ref{sec:3}). A central star embedded within such a dusty gas cloud might stay undetected, if it is sufficiently small and either very hot or very cool, but the presence of such a compact central source would be irrelevant for the dynamics of the observable gas. The gas dynamics at pericenter are dominated by the gravitational force of the black hole and possibly affected by the ambient medium.

In this paper we report on new near-infrared observations of both the ionized gaseous component and the dusty component of G2, obtained in 2015 and 2016 with the SINFONI integral field spectrograph and the NACO imager at the ESO~VLT. First, we summarize the observations and describe the data reduction and processing procedures, as well as the modeling techniques~(Sec.~\ref{sec:2}). We then analyze the post-pericenter evolution of these two components in the context of our previous observations, which allows us to draw the most comprehensive picture of G2 to date~(Sec.~\ref{sec:3}). We also determine a revised orbit solution by using an improved numerical approach. Finally, we discuss how observations of G2 can constrain properties of the accretion flow, how it is related to the precursor source G1 and the trailing tail emission, as well the implications for its nature, origin and fate~(Sec.~\ref{sec:4}).

\section{Observations \& Methods}
\label{sec:2}

We have obtained new, deep near-infrared observations of the Galactic Center at the ESO~VLT with SINFONI \citep{2003Msngr.113...17E,2004Msngr.117...17B} in April/May 2015 and April/July 2016, as well as with NACO \citep{1998SPIE.3354..606L,1998SPIE.3353..508R} in July/September 2015 and April 2016.

\newpage

\subsection{SINFONI data reduction}
\label{subsec:2.1}

As previously \citep[e.g.][]{2015ApJ...798..111P}, we have used the narrow-field camera (${25~\mathrm{mas~px^{-1}}}$) of SINFONI with the H+K grating and the adaptive optics in natural guide star mode or, if available, the laser guide star, in combination with an object-sky-object observing pattern and a detector integration time of ${600~\mathrm{s}}$. By using a quadratic dither pattern, the total field of view is extended from originally ${0.8''\times0.8''}$ to ${1.2''\times1.2''}$. The central region containing Sgr~A* and G2 is still covered at every dither position and ultimately sampled at ${12.5~\mathrm{mas~px^{-1}}}$ with a spectral resolution of about $1500$.

We use the data reduction pipeline \textit{SPRED} \citep{Abuter:2006kc} to perform flat-fielding and sky subtraction, wavelength and telluric calibration, as well as a bad pixel and distortion correction. We also refine the default wavelength calibration based on the observed atmospheric OH lines and we select only the high-quality data for which the FWHM of the star S2 is smaller than ${7.5~\mathrm{px}}$ at a wavelength of ${2.2~\mu\mathrm{m}}$. The total integration time spent on source in 2015 is ${810~\mathrm{min}}$, out of which ${600~\mathrm{min}}$ pass this quality cut. In 2016 the total is ${520~\mathrm{min}}$, out of which we are able to use ${400~\mathrm{min}}$. Due to poor weather conditions these exposures are not as deep as those obtained in previous years. In a final step we reconstruct a single combined data cube for each year, from which it is possible to extract channel maps, regionalized spectra and position-velocity (pv) diagrams to isolate the recombination line emission from G2. The orbital evolution of G2 is slow enough so that combining observations from a few consecutive months does not skew the data, but on the contrary maximizes the signal to noise ratio. Adding the two data cubes from 2015 and 2016 to our existing data set increases the total number of epochs to $10$ since 2004.

For each data cube we manually choose an extraction map that defines source and background pixels. The line emission from G2 has never been strongly confused with other sources due to G2's trajectory and in particular its large radial velocity, but to attain a clean signal it is necessary to choose the source- and background regions with care and remove atmospheric and stellar emission. Atmospheric features make individual spectral channels appear brighter or fainter, whereas stellar (continuum) emission leads to brightness variations across all spectral channels at a given position. Subtracting the appropriate median value at each position and then subtracting the average background in each channel removes these artifacts reliably. Even so, the choice of source and background pixels remains the most likely source of a systematic uncertainty. For the astrometric and photometric calibration we use a changing set of relatively isolated stars, the motions and (K-band) magnitudes of which are well know from the long-time monitoring of stellar motions in the Galactic Center \citep{2009ApJ...692.1075G}.

If taking a certain on-sky trajectory as given is justified, the complexity of the spectro-spatial information in a three-dimensional data cube can be reduced in an optimal way into a two-dimensional pv-diagram, to visualize the low surface-brightness emission close to the detection limit. In practice, we construct pv-diagrams of G2 by placing virtual slits (${8~\mathrm{px}}$ wide) along the trajectory of the best-fit orbit and aligning the resulting spectra as a function of distance along this path. To increase the signal to noise ratio further, we create noise-weighted, coadded diagrams using both the Br$\gamma$ and He$\RN{1}$ lines.

\subsection{NACO image processing}
\label{subsec:2.2}

Because of a malfunction of NACO's camera wheel in 2015 we have used the L' filter (${\lambda=3.80\pm0.31~\mu\mathrm{m}}$) in combination with the S13 camera, which according to specification only supports a wavelength range of ${1.0~\mu\mathrm{m}}$ to ${2.5~\mu\mathrm{m}}$. A longer detector integration time and total exposure time than usual were necessary to compensate for the reduced sensitivity, but the smaller pixel scale (${13~\mathrm{mas~px^{-1}}}$) and the low level of image distortion of the S13 camera were advantageous in comparison to the normally used L27 camera. In 2016 we were able to use the standard configuration again.

Because of another issue with the readout electronics, every 8th column of one detector quadrant showed no signal. This problem is easily mitigated by replacing the missing values in every frame with interpolated values based on the neighboring detector columns, since the point spread function (PSF) is well sampled.

After this correction we apply standard flat-field and bad pixel corrections. A sky image is constructed individually for each object image by calculating the median image of several other randomly offset images taken close in time. Any images of exceptionally bad quality are rejected, for example those affected by a thin cloud cover. We finally use the position of a common bright star as a reference position to combine the remaining images into a mosaic image. Despite the different instrumental setups, the mosaic images from 2015 and 2016 reach a similar depth in the central region around Sgr~A*.

We extract an empirical PSF from the mosaic images using the tool StarFinder \citep{2000A&AS..147..335D} and use this PSF in a Lucy-Richardson algorithm to deconvolve the images to a moderate depth. From the resulting images we measure the positions and instrumental magnitudes of several stars that are suitable as astrometric and photometric reference sources \citep{2009ApJ...692.1075G}. We are thus able to subtract from the images the bright B-stars that are at the time of observation confused with G2 in L'-band (in particular S2, S19 and S31), assuming a constant K-L' color for all stars of this common spectral type. Apart from Sgr~A* there are no unaccounted sources in the central region brighter than about 17th to 18th magnitude in K-band.

The subtraction of the infrared source Sgr~A* is not as straightforward, because it is intrinsically variable on a time scale of tens of minutes during flaring events that occur sporadically. However, only the 2015 data shows significant emission at the position of Sgr~A* and at this time G2 is already spatially distinguishable from Sgr~A* due to its orbital motion, even though the images of the two sources still partly overlap in projection. This allows us to measure the L'-band light curve of Sgr~A* by performing small-aperture photometry on the series of single images and measure accurately the average brightness that we have to subtract.

\begin{figure*}
\centering
\includegraphics[width=\linewidth]{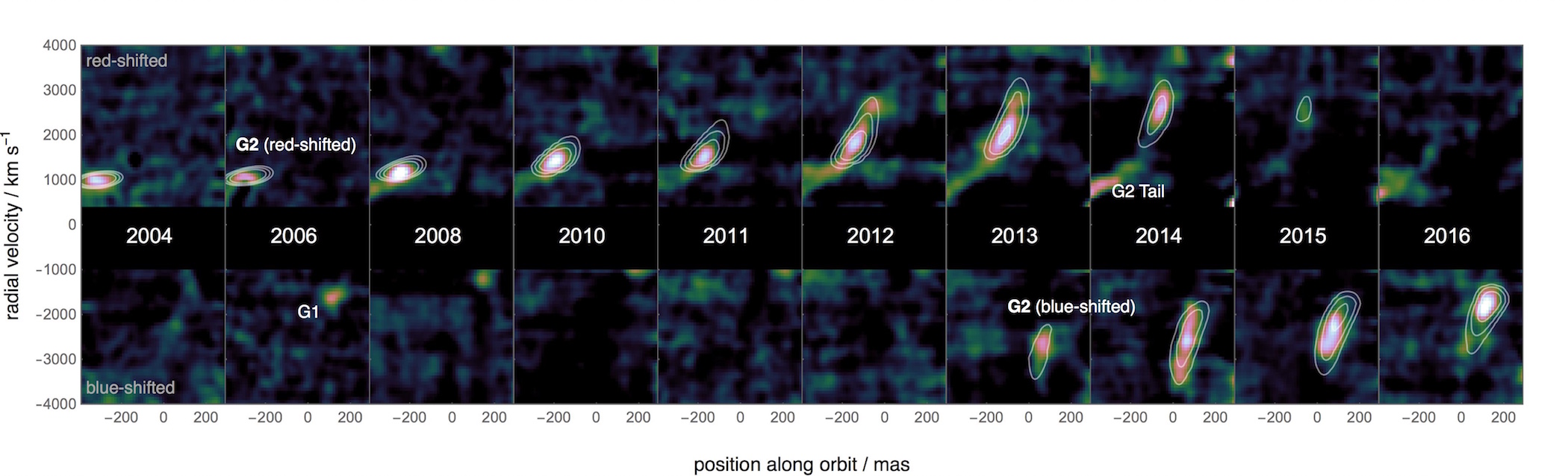}
\caption{Position-velocity diagrams of the coadded Br$\gamma$ and He$\RN{1}$ line emission from G2 between 2004 and 2016 (shown in color, noise-normalized), extracted along the updated orbit and superimposed with a projection of the best-fit test-particle cloud (shown in contours). The region around zero velocity is masked because it is affected by background emission and stellar absorption features. The placement of the slit is optimized for G2 but also partially covers G1, which appears as a blue-shifted source at early times, and the tail of G2, which appears close to zero velocity at later times (but see fig.~\ref{fig:9}). In these diagrams, a velocity gradient appears as a tilt (see also fig.~\ref{fig:4}).}
\label{fig:1}
\end{figure*}

\subsection{Orbit Fitting}
\label{subsec:2.3}

We let G2 be represented by an ensemble of independent massless particles to fit the orbital evolution of the Br$\gamma$-emitting gas cloud over a time span of $12$ years, from well before to right after the pericenter passage. Such a test-particle cloud has already proven to be a good description of past observations \citep{2012Natur.481...51G,2013ApJ...763...78G,2013ApJ...774...44G,2015ApJ...798..111P} and has several benefits as a model. A test-particle simulation is far less computationally demanding than a hydrodynamic simulation, but still offers a way to account for other external forces in addition to gravity. More importantly, it allows a direct comparison to the observations in the (position-velocity) space of the data.

A possible interaction with the ambient gas in the accretion zone is accounted for by including a distance- and velocity-dependent drag force on the particles in the simulation that has the form of a ram pressure:
\begin{equation}
\bm{a}_D=-c_D\left(\frac{r}{\mathrm{as}}\right)^\alpha\left(\frac{v}{\mathrm{as/yr}}\right)^{2}\frac{\bm{v}}{v}\frac{\mathrm{as}}{\mathrm{yr}^2}
\label{eq:1}
\end{equation}
The radial dependency is characterized by the exponent $\alpha$, which has a value of ${\alpha=-1.5}$ in a spherical Bondi accretion model \citep{2003ApJ...598..301Y}, but has been measured from X-ray observations to be somewhat closer to ${\alpha=-0.5}$ in the Galactic Center \citep{2013Sci...341..981W}. However, G2 probes the accretion zone at much smaller distances than the X-ray observations can and we therefore consider both extreme cases. The free parameter $c_D$ determines the strength of the drag force and is assumed to be a constant non-negative coefficient. In the absence of a drag force (${c_D=0}$), a particle in the point mass potential generated by the central black hole moves in a fixed orbital plane, and on an ellipse if it is bound. Such a Keplerian orbit is fully described by six parameters: The semi-major axis $a$ and the eccentricity $e$, which describe the shape of the orbit, the three angles inclination $i$, longitude of ascending node $\Omega$ and argument of pericenter $\omega$, which describe the orientation of the orbital plane, as well as the time of pericenter $t_p$. In the presence of a drag force (${c_D>0}$), the orientation of the orbital plane is unchanged, but the particles experience an instantaneous deceleration oriented against their direction of motion and deviate from elliptical orbits. Because the trajectory of G2 is almost radial and the ambient density increases exponentially towards the center, the effect of the drag force is concentrated around the time of pericenter.

We use the \textit{MultiNest} implementation \citep{2009MNRAS.398.1601F} of the nested sampling algorithm \citep{2004AIPC..735..395S} to calculate the marginalized likelihood\footnote{The (fully) marginalized likelihood of a model is the expected value of the likelihood over the prior volume. It is also commonly known as the Bayesian evidence \citep{MacKay:2003wc}.} to be able to compare different models and at the same time sample from the posterior probability distribution of the model parameters. To evaluate the likelihood once for a certain set of orbital parameters, we run a complete simulation starting at epoch $t_0$ (well before the first observations), using $10^3$ independent test particles that are initially placed on the given orbit and then perturbed randomly in position, where the offsets are sampled randomly from a spherically symmetric normal distribution. Both the starting time and the size of the initial particle distribution are included as two additional parameters in the model. The number of test particles should be chosen as large as possible to limit the noisiness of the likelihood, which is introduced by the inherent randomness of the initial conditions, and diminishes the efficiency of the sampling process.

The particle orbits are numerically integrated using the \textit{REBOUND} multi-purpose n-body code \citep{2012A&A...537A.128R} and specifically the \textit{whfast} integrator \citep{2015MNRAS.452..376R} and the high-accuracy \textit{ias15} integrator \citep{2015MNRAS.446.1424R}. The former algorithm is optimized for Keplerian orbits, whereas the latter algorithm adaptively reduces the time step of the simulation to resolve close encounters of particles with the black hole, during which they can experience very strong accelerations. We assume a black hole mass of ${4.31\times10^6~M_\odot}$, a distance to the Galactic Center of ${8.33~\mathrm{kpc}}$ and uninformative priors for all sampled parameters, which are chosen either uniform or uniform in log-space in case of the scaling parameters.

Upon reaching each epoch of observation as the orbit integration proceeds, the test-particle cloud is convolved with a Gaussian model PSF to match the resolution of the observations (${80~\mathrm{mas}}$ and ${120~\mathrm{km~s^{-1}}}$ FWHM along the spatial and spectral dimensions, respectively) and then compared directly to the relevant volume of the data cube around the Br$\gamma$ line, which is stored as a background-subtracted table of source pixels, assuming zero-mean constant Gaussian noise to define the likelihood function (i.e. a background-dominated regime).

A major advantage of this specific fitting technique is that the entire available spectro-spatial information is made use of, which for example includes velocity gradients that manifest themselves as changes of radial velocity across the cloud. In this way projection effects are naturally accounted for as well, which can have a strong impact on the appearance of G2 in the observations, depending on the orientation of the orbital plane. For instance, the peak or centroid of the cloud emission does not in general correspond to that of the center of mass if the cloud is extended, along any dimension of the data, as might be implicitly assumed. This effect could cause a bias when only these data points were used for fitting the orbit and not the data cubes.

\section{Results}
\label{sec:3}

\subsection{Gas emission}
\label{subsec:3.1}

\begin{figure*}
\centering
\includegraphics[width=0.95\linewidth]{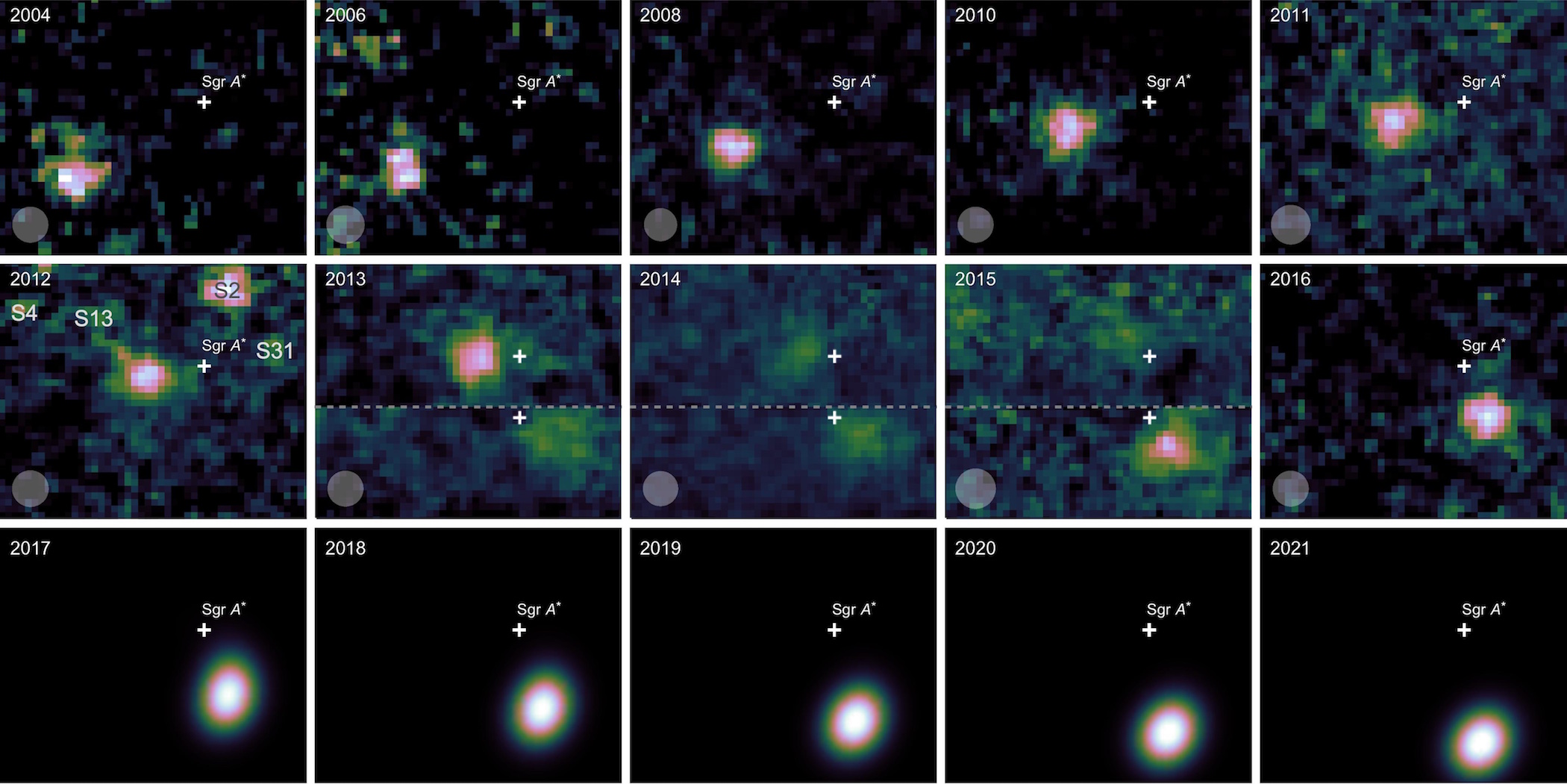}
\caption{Top Rows: Velocity-integrated maps of the Br$\gamma$ line emission from G2 between 2004 and 2016 (shown in color, flux-normalized). In 2013, 2014 and 2015 the blue- and red-shifted components are extracted from different regions and shown separately. The dimensions of each image are 0.6'' by 0.5'' (North is up and East is left) and the approximate size of the PSF (FWHM) is indicated by the opaque circle. Bottom row: Predicted appearance of G2 over the next five years according to the test-particle simulation.}
\label{fig:2}
\end{figure*}

\begin{figure*}
\centering
\includegraphics[width=0.4\linewidth]{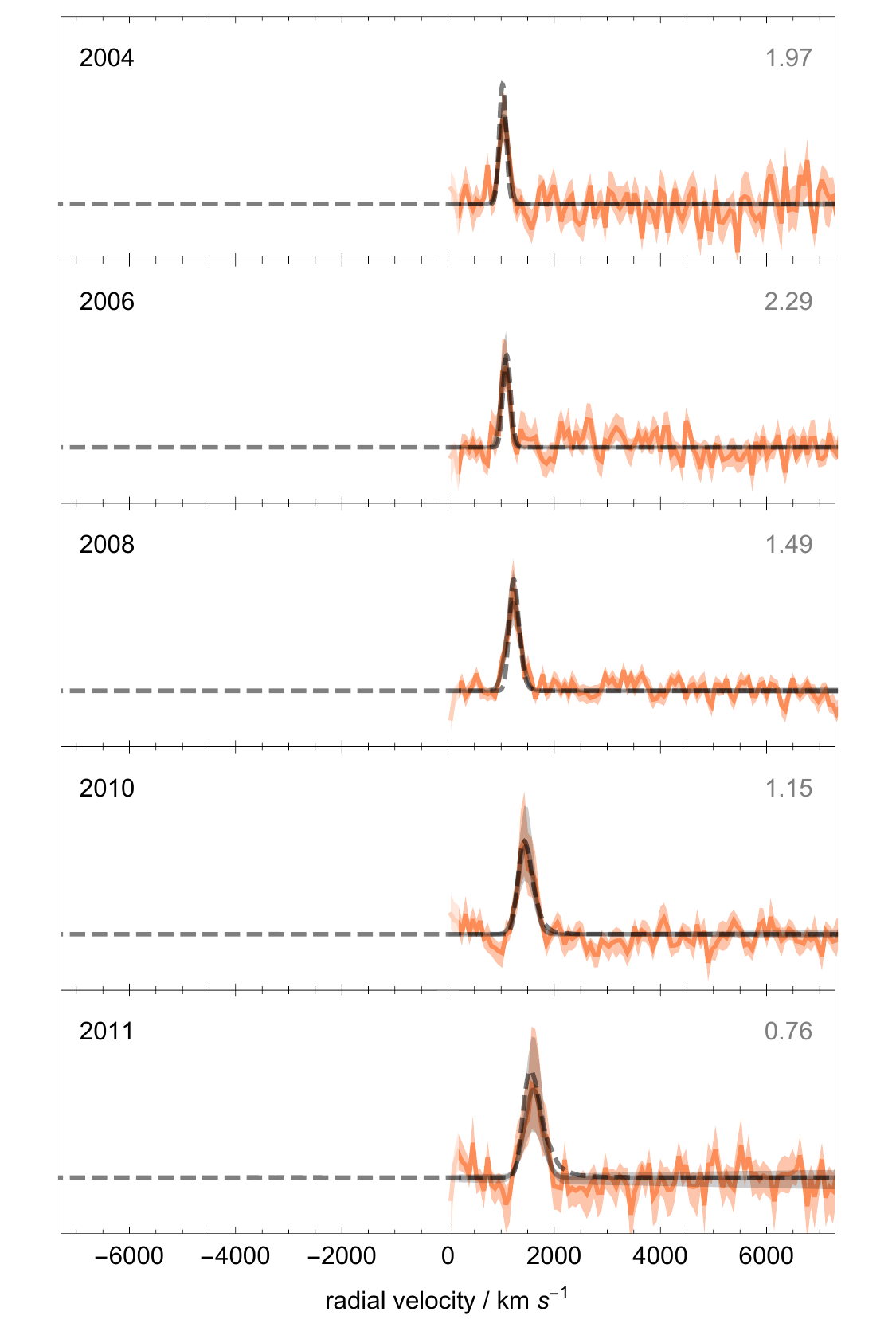}
\includegraphics[width=0.4\linewidth]{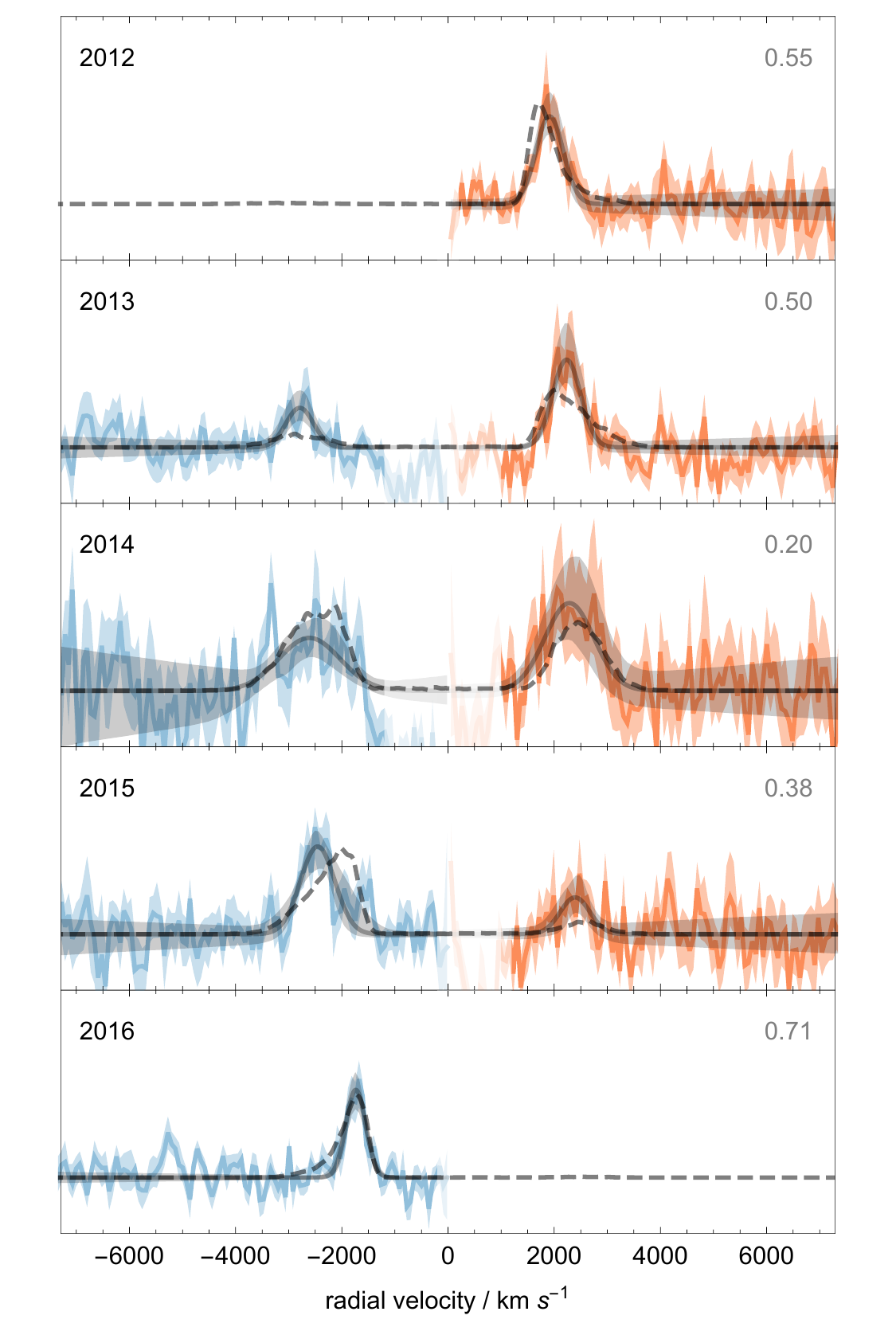}
\caption{Radial velocity distribution of the Br$\gamma$ line emission from G2 between 2004 and 2016 (colored lines, peak-normalized), superimposed with that of the best-fit test-particle cloud (dashed line, scaled to the mean total luminosity). In 2014 and 2015 the blue- and red-shifted components are extracted from different regions but shown together in this graphic. Also shown is the fit of a Gaussian profile to the Br$\gamma$ lines (solid black lines with error bands). The region around zero velocity is excluded from this fit, because it is affected by background emission and stellar absorption features. The relative peak value from the fit is indicated by the number in the top right corner.}
\label{fig:3}
\end{figure*}

\begin{figure*}
\centering
\includegraphics[width=0.95\linewidth]{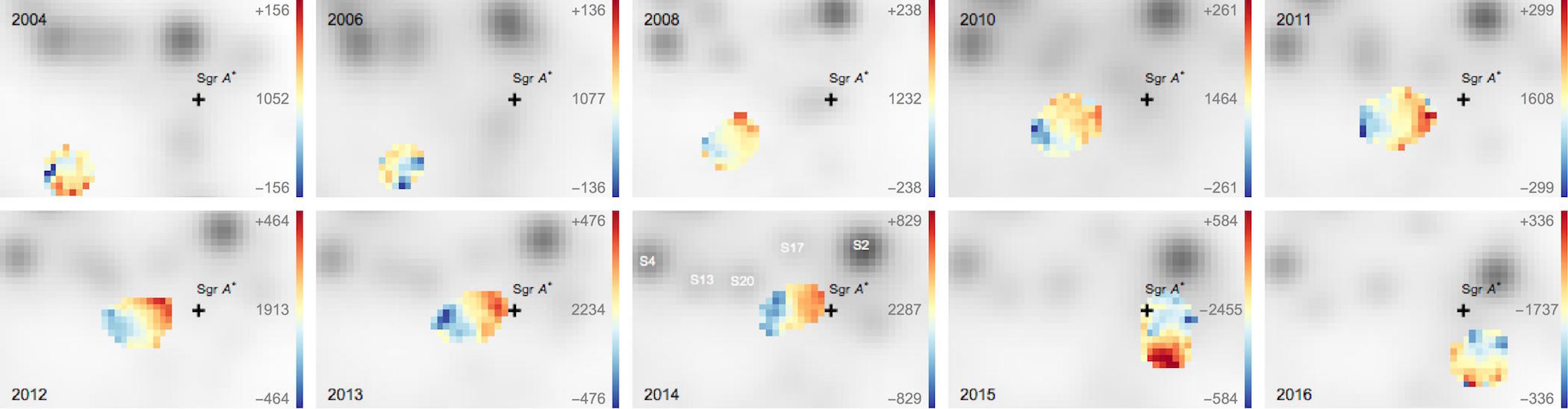}
\caption{Radial velocity maps of the Br$\gamma$ line emission from G2 between 2004 and 2016, where a change in color indicates a velocity gradient across the gas cloud. The color scale in each map is adjusted according to the measured bulk velocity and line width (Fig.~\ref{fig:3}). A spatially resolved velocity gradient is most clearly detected after 2010. This is also evident from the pv-diagrams (Fig.~\ref{fig:1}), but can be seen in this visualization without making any assumptions about the cloud trajectory. For reference, a K-band continuum image is shown in the background, on which nearby stars are marked. The dimensions of each image are 0.6'' by 0.4'' (North is up and East is left).}
\label{fig:4}
\end{figure*}

\begin{figure}
\centering
\includegraphics[width=0.95\linewidth]{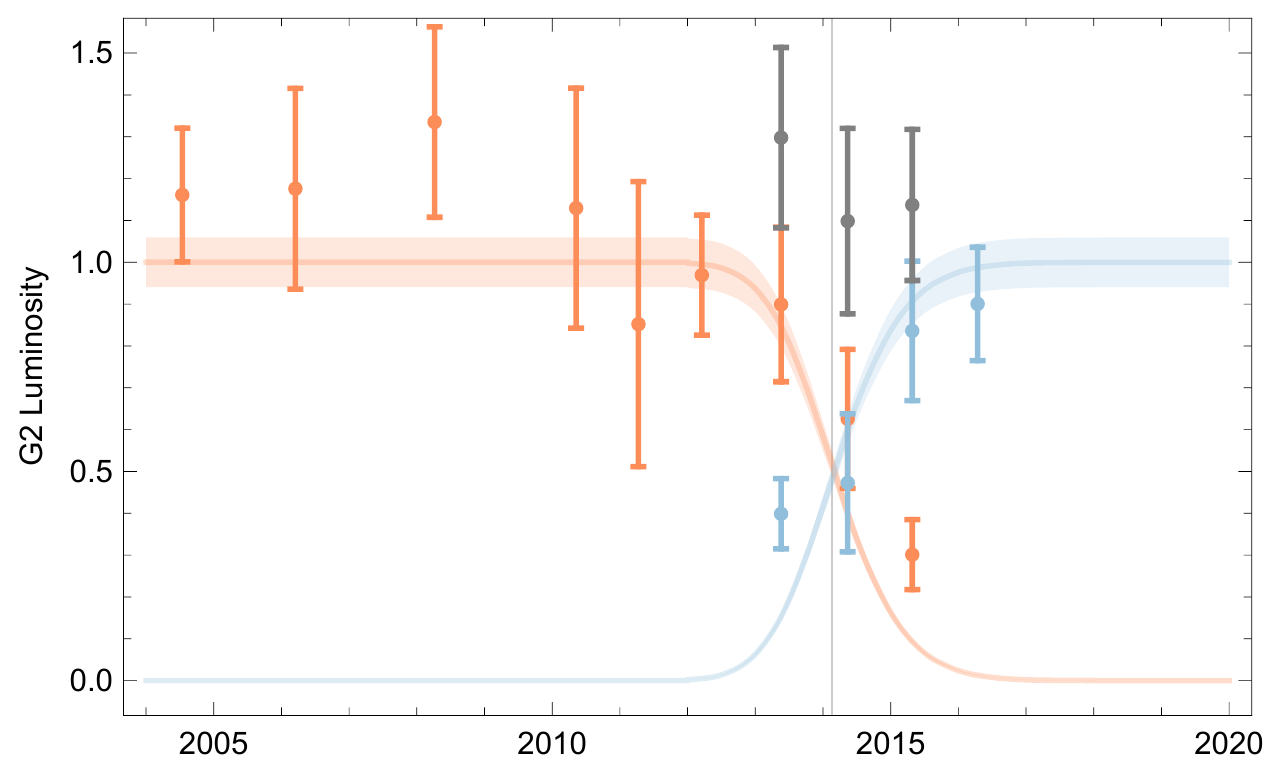}
\caption{The luminosity of the Br$\gamma$ line emission from G2 as a function of time (colored points). Also shown is the relative amount of red- and blue-shifted material, based on the best-fit test-particle simulation (colored lines, scaled to the mean total luminosity). If red- and blue-shifted components are detected simultaneously, the total luminosity is shown as an additional data point (in black). A strict comparison of the data points to the curves is only valid if the observed emission is a direct proxy for mass throughout the evolution. The closest approach of the center of mass, by definition at the time of pericenter (vertical line), happens shortly before half of the mass has changed from being red- to blue-shifted. The entire pericenter passage is a prolonged process stretching over several months from 2013 to 2016. The error bars stem from resampling the selected source and background pixels and the error bands reflect the uncertainty of the mean luminosity.}
\label{fig:5}
\end{figure}

\begin{figure*}
\centering
\includegraphics[width=0.8\linewidth]{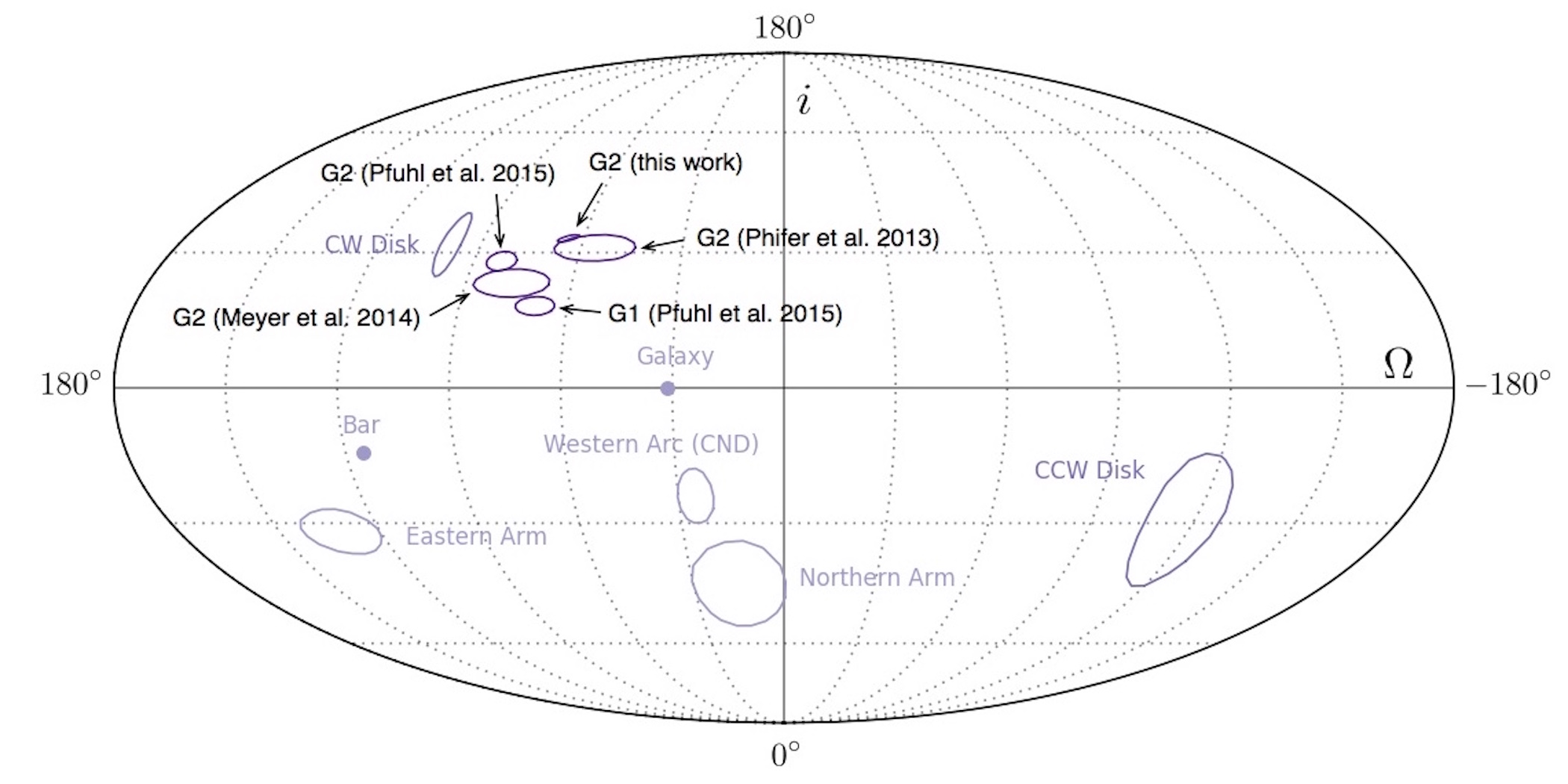}
\caption{Estimates of the orientation of G2's orbital angular momentum vector in comparison to other orbit solutions from the literature \citep{2013ApJ...773L..13P,2014IAUS..303..264M,2015ApJ...798..111P} and the orientations of other planar structures in the Galactic Center environment \citep[][Tab.~1 and references therein]{2010RvMP...82.3121G}.}
\label{fig:6}
\end{figure*}

\begin{figure}
\centering
\includegraphics[width=0.95\linewidth]{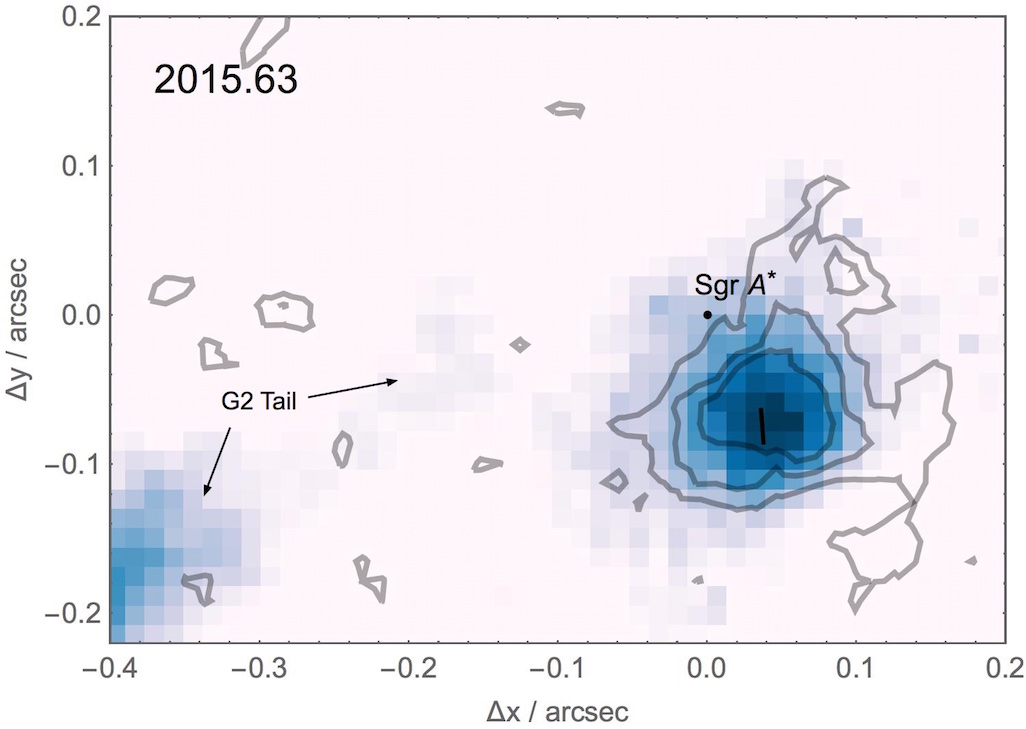}\\
\includegraphics[width=0.95\linewidth]{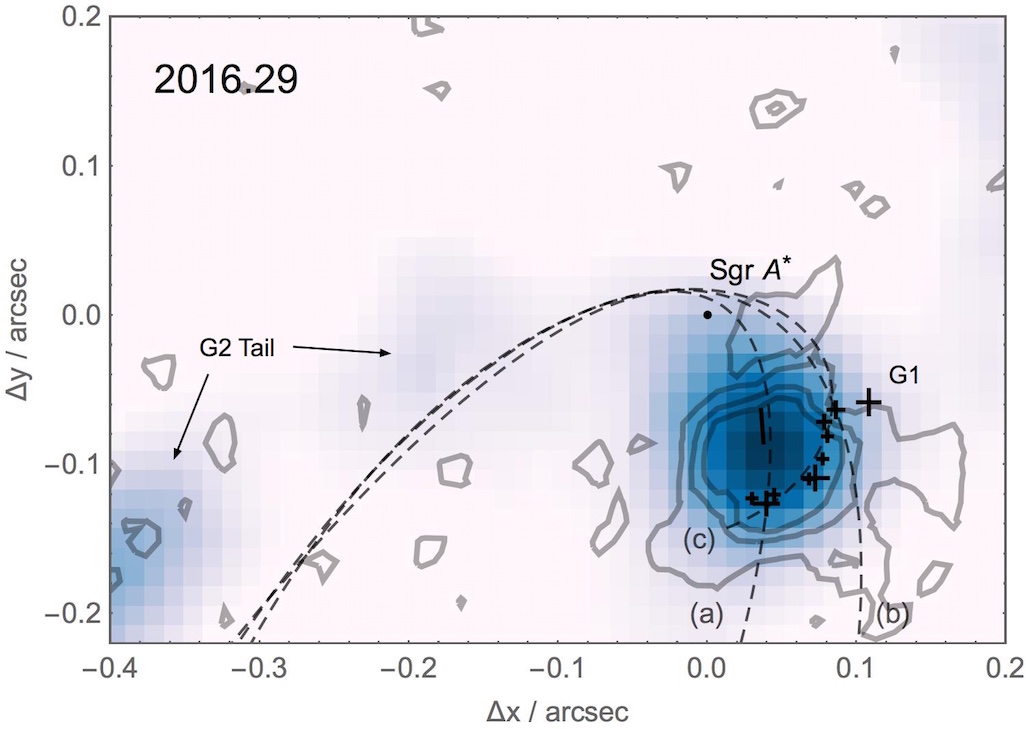}
\caption{Images of the L' emission from G2 in 2015 and 2016 (shown in color), superimposed with the blue-shifted Br$\gamma$ line emission at similar times (shown in contours), which appears at the same location. All stellar sources in the region and any significant emission from Sgr~A* have been subtracted~(Sec.~\ref{subsec:2.2}). For ease of comparison, the on-sky motion of the centroid between the two epochs is marked (solid line). The faint L' emission trailing G2 might be part of the larger tail structure (Sec.~\ref{subsec:4.3}). In the bottom panel, the on-sky trajectory of (a) the updated best-fit orbit of G2, (b) the orbit of G2 from \citet{2015ApJ...798..111P} and (c) their combined orbit of G2/G1 is overplotted (dashed lines, see annotations). The indicated positions of G1 (crosses) were measured between 2003 and 2010.}
\label{fig:7}
\end{figure}

The new observations from 2015 and 2016 reveal the continued evolution of G2 immediately after its closest approach to Sgr~A*. Figure~\ref{fig:1} shows how the complete pericenter passage has unfolded over the last 12 years in a time series of 10 pv-diagrams (Sec~\ref{subsec:2.1}), which we extracted along the updated best-fit orbit (Table~\ref{tab:1}). Figures~\ref{fig:2}, \ref{fig:3} and \ref{fig:4} show corresponding sets of integrated maps, velocity distributions and velocity maps of the gas emission, which are complementary visualizations of the same underlying three-dimensional data.

The gas emission has changed from being still partially red-shifted in 2015 to being entirely blue-shifted by 2016. This confirms that as predicted the bulk of cloud material has passed the black hole in 2014 at high velocity, assuming the emission is a valid proxy for gas mass throughout the evolution. The blue-shifted Br$\gamma$ line is detected at a radial velocity of ${-2400~\mathrm{km~s^{-1}}}$ in 2015 and redetected at ${-1700~\mathrm{km~s^{-1}}}$ in 2016, with a respective line width of ${320~\mathrm{km~s^{-1}}}$ and ${190~\mathrm{km~s^{-1}}}$.

The total Br$\gamma$ luminosity of G2 continues to be comparable to previous measurements, while the ratio of blue- to red-shifted emission has changed rapidly over just a few years (Fig.~\ref{fig:5}). The fraction of blue-shifted emission increased from 30\% to 70\% between 2013 and 2015, before reaching 100\% by 2016. During this time the total luminosity has remained constant and at most could have weakly increased during pericenter.

The leading gas in 2015 must be spatially extended as it continues to show a pronounced velocity gradient due to tidal shear that is spatially well-resolved (Fig.~\ref{fig:1} \&~\ref{fig:4}). This extent might not be obvious from just a map of the integrated line emission (Fig.~\ref{fig:2}), since the motion of G2 is mostly seen in radial velocity and only one other on-sky coordinate, due to the almost edge-on orientation of the orbit. A velocity gradient on the order of ${10~\mathrm{km~s^{-1}~mas^{-1}}}$ is still apparent in 2016, even though the gas cloud now appears more compact again in velocity space (Fig.~\ref{fig:1}). This is expected if the cloud is purely interacting gravitationally with the black hole and hardly at all with the ambient gas. Over the next few years the appearance of the cloud will hardly change, assuming the test-particle model stays valid (Fig.~\ref{fig:2}, bottom row).

The post-pericenter Br$\gamma$ data (Sec.~\ref{subsec:2.1}) and the improved methods of analysis (Sec.~\ref{subsec:2.3}) provide a new set of orbital parameters that typically have half of their previous statistical uncertainty or less. The updated best-fit parameters broadly agree with those found by \citet{2015ApJ...798..111P} to within these uncertainties, except that we find a different orientation of the orbital plane (${\Delta\Omega\approx20^\circ}$). The estimates of all orbital parameters are summarized in table~\ref{tab:1} and their eight-dimensional posterior distribution is presented in more detail in the appendix (Fig.~\ref{fig:11}). The angular offset between the orientation of G2's orbital plane and that of the clockwise stellar disk amounts to~$30^\circ$ (Fig.~\ref{fig:6}), while the apocenter of G2 at about~$1''$ projected distance from Sgr~A* still falls within the inner disk region \citep{2006ApJ...643.1011P,2009ApJ...697.1741B,2014ApJ...783..131Y}. The main characteristic of the orbit is unchanged, which is the almost radial trajectory. The best-fit pericenter distance is ${18.4\pm0.9}$ light hours or ${1560\pm70~R_S}$. In terms of a marginalized likelihood ratio (i.e. the Bayesian evidence ratio), none of the models that we tested in which G2 is affected by an additional drag force~(Sec.~\ref{subsec:2.3}) is significantly favored over the Keplerian model.

However, none of the fits can accurately describe the pericenter observations in detail (Fig.~\ref{fig:1} \&~\ref{fig:3}). The gas seems to overshoot at the beginning of the pericenter passage, when in 2013 a substantial fraction of the emission already appears on the blue-shifted side, while as much gas should not have had time to arrive there. The gas also lags behind at the end, when in 2015 a substantial fraction of emission stays present on the red-shifted side, while the gas should have almost finished passing to the other side.

\begin{deluxetable}{lcc}
\tablecaption{Summary of the model parameters, which describe a test-particle cloud on a Keplerian orbit that fits the observed evolution of G2 in Br$\gamma$ emission between 2004 and 2016.}
\tablehead{\colhead{Model Parameter} & \colhead{Value\tablenotemark{a}}}
\startdata 
Semi-Major Axis \\
$a$ (as) & $0.96\pm0.07$\\
Eccentricity \\
$e$ & $0.983\pm0.002$ \\
Inclination \\
$i$ (deg) & $123\pm1$ \\
Longitude of Ascending Node \\
$\Omega$ (deg) & $65\pm3$ \\
Argument of Pericenter \\
$\omega$ (deg) & $92\pm2$ \\
Time of Pericenter \\
$t_p$ (yr) & $2014.18\pm0.09$ \\
Initial Cloud Size (FWHM) \\
$w_r$ (mas) & $50\pm6$\\
Starting Time \\
$t_0$ (yr) & $1975\pm8$\\
\hline Orbital Period \\
$P$ (yr)& $346\pm37$ \\
\enddata
\tablenotetext{a}{Given is the median value of the marginal posterior distribution of each parameter (see fig.~\ref{fig:11}).}
\label{tab:1}
\end{deluxetable}

\subsection{Dust emission}
\label{subsec:3.2}

In the cleaned L' images, from which confusing sources have been removed~(Sec.~\ref{subsec:2.2}), we clearly detect a residual source in both 2015 and 2016 that we attribute to thermal continuum emission from dust associated with G2 (Fig.~\ref{fig:7}). The location of this source is in very good agreement with that of the blue-shifted Br$\gamma$ emission at similar times. While the Br$\gamma$ emission is resolved in position-velocity space (Sec.~\ref{subsec:3.1}), the L' emission is unresolved in our images. That is to say it is not significantly elongated with respect to the estimated PSF, which has a FWHM on the order of ${120~\mathrm{mas}}$. However, due to the high background emission in this crowded region, we are only able to observe the very concentrated part of the dust distribution, which could intrinsically still be very similar to the gas distribution. We estimate the L' magnitude of G2 to be ${13.8\pm0.5}$ in 2015 and ${13.2\pm0.5}$ in 2016, which is consistent with a constant or possibly weakly increased luminosity compared to previous measurements before and around the time of pericenter \citep{2014ApJ...796L...8W}.

\section{Discussion}
\label{sec:4}

\subsection{G2 as a probe of the accretion zone}
\label{subsec:4.1}

\begin{figure}
\centering
\includegraphics[width=0.95\linewidth]{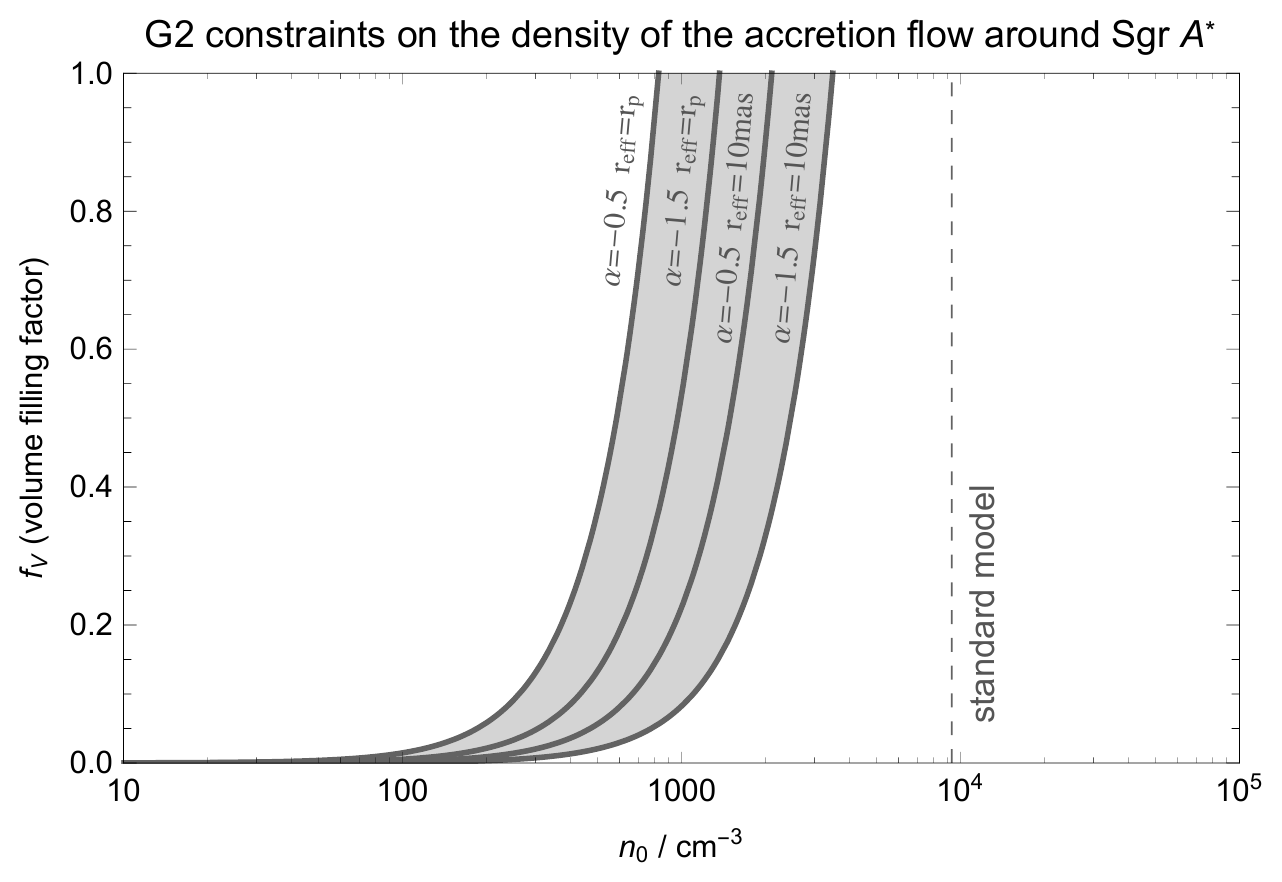}
\caption{The upper limit on the drag force acting on G2 constrains the density of the ambient gas at the pericenter distance and the volume filling factor of G2 to the shaded region, assuming G2 is a spherical cloud with an effective radius between ${10~\mathrm{mas}}$ and ${16~\mathrm{mas}}$ ${(\approx r_p)}$ that is moving through a stationary and homogeneous atmosphere. The assumed slopes of the ambient density profile ${n(r)=n_0\times(r/r_p)^\alpha}$, ${\alpha=-1.5}$ and ${\alpha=-0.5}$, yield similar exclusion regions in this graphic. The density indicated by the vertical line is derived from the accretion model by \citet{2003ApJ...598..301Y} and \citet{2006ApJ...640..319X} that was used by \citet{2012Natur.481...51G} and others \citep[e.g.][]{2012ApJ...750...58B,2012ApJ...755..155S} to predict the pericenter and post-pericenter evolution of G2.}
\label{fig:8}
\end{figure}

The non-detection of a strong drag force on G2 during its pericenter passage (but see Sec.~\ref{subsec:3.1}) allows us to constrain internal properties of the gas cloud, as well as external properties of the ambient gas in the inner accretion zone. To first approximation, the drag force acting on G2 as it moves along on its orbit, assuming a stationary and homogeneous atmosphere, is expected to have the form of a ram pressure:
\begin{equation}
\bm{F}_D=M_\mathrm{cl}\bm{a}_D\approx-\frac{1}{2}A_\mathrm{eff}\rho v^2\frac{\bm{v}}{v}
\label{eq:2}
\end{equation}
The accretion flow is fully ionized and charge neutral, so the ambient mass density $\rho$ converts into an electron number density ${n=\rho/m_p}$. We parameterize the density profile $n(r)$ as a radial power law, normalized to the density at the pericenter distance (${r_p\approx1500~R_s}$).
\begin{equation}
n(r)=n_0\left(\frac{r}{r_p}\right)^\alpha
\end{equation}
G2 was estimated to have a mass of ${M_\mathrm{cl}\approx3~M_\Earth\sqrt{f_v}}$, $f_V$ being the volume filling factor, and an effective cross section of ${A_\mathrm{eff}=\pi r_\mathrm{eff}^2}$ with ${r_\mathrm{eff}\approx15~\mathrm{mas}}$ in 2011 \citep{2012Natur.481...51G,2014ApJ...783...31S}. Additionally, we find from the test particle simulations that a reasonable minimum size of a cloud resembling G2 at pericenter is ${10~\mathrm{mas}}$. Based on these assumptions we are able to relate the numeric drag force coefficient $c_D$ to the ambient density $n_0$ and the volume filling factor $f_V$, by comparing equations~\ref{eq:1} and~\ref{eq:2}. In all models that we tested, the parameter $c_D$ initially converges towards the edge of the prior volume at zero (assuming a strictly resistive drag force). The final best-fit value is non-zero but small, and the resulting drag force has a negligible impact on the orbital evolution. However, this value represents an upper limit on $c_D$ that yields a constraint on the ratio ${n_0/\sqrt{f_V}}$ (Fig.~\ref{fig:8}). In particular, it must be ${n_0\lesssim10^3~\mathrm{cm^{-3}}}$ for filling factors close to unity. The assumed slope of the density profile has only a minor influence on the order of magnitude of this limit.

X-ray observations constrain the ambient density to about ${10^2~\mathrm{cm^{-3}}}$ at the scale of the Bondi radius (${10^5~R_s}$), assuming Bremsstrahlung is the origin of the observed gas emission \citep[e.g.][]{2003ApJ...591..891B}. At a scale of just a few $R_s$, radio observations constrain the density to about ${10^7~\mathrm{cm^{-3}}}$, via measurements of the polarization (i.e. Faraday Rotation) and the size of the emission from Sgr~A* \citep[e.g.][]{2000ApJ...545..842Q,2003ApJ...588..331B,2007ApJ...654L..57M,2008Natur.455...78D}. Both of these constraints can be brought into agreement if the density profile is a smooth power law with a slope of ${\alpha\approx-1}$. In comparison to this prediction, our upper limit at around ${10^3~R_s}$ is smaller by roughly an order of magnitude. If the density of the inner accretion zone is indeed this small, it would imply a broken power law with a slope that is shallower at larger radii (${\alpha\approx-0.5}$) and steeper at smaller radii (${\alpha\approx-1.4}$). It would also explain the non-detection of excess radio or X-ray emission above the quiescent level of Sgr~A*, because the expected observable emission from a shock front would not exceed this threshold \citep[following][]{2012Natur.481...51G}. A possible co-rotation of the accretion zone could additionally reduce the velocity of the gas cloud relative to the local gas (i.e. the Mach number), to a similar effect \citep[e.g.][]{2012ApJ...757L..20N}. A lower ambient density would also allow proportionally larger cloud sizes, without the need for additional mechanisms of (e.g. magnetic) confinement \citep{2014ApJ...783...31S,2015MNRAS.449....2M}.

The remaining discrepancy between the test-particle model and the observations (see Sec.~\ref{subsec:3.1}) might be explained by additional material behind or in front of the cloud, or by a more complex interaction between G2 and the accretion flow than the drag force we assumed in our model. For example, the shape of G2 might also not be spherical or there might be a rotation of the accretion flow \citep{2013MNRAS.433.2165S,2016MNRAS.455.2187M}, an in- or outflow \citep{2016arXiv160202760M} or dynamically important magnetic fields \citep{2015MNRAS.449....2M}. These effects could alter the strength and also the direction of a drag force. It can also not be excluded that the properties of the emission process are variable. In addition, the data analysis is most susceptible to systematic uncertainties around the time of pericenter, due to the low surface brightness of the observed emission. The test-particle model still performs surprisingly well at describing the overall evolution of the gas cloud between 2004 and 2016.

\subsection{The relationship between G2 and G1}
\label{subsec:4.2}

A relationship between G2 and G1 was first proposed and investigated by \citet{2015ApJ...798..111P} and later studied in more detail by \citet{2016MNRAS.455.2187M} and \citet{2016arXiv160202760M}, but challenged by \citet{2015AAS...22510207S} on the basis that their orbital elements are inconsistent. The G1 object, which was first recognized in observations by \citet{2004A&A...417L..15C,2004A&A...424L..21C,2005A&A...439L...9C} and \citet{2005ApJ...635.1087G}, is a source apparently much alike G2 that passed the black hole slightly more than a decade earlier and was only observed after its pericenter passage. It started to fade soon afterwards, so that no follow-up detections could be verified.

The post-pericenter trajectory of G2 roughly matches the one of G1 (Fig.~\ref{fig:7}), but goes on in a more southern direction. However, the orbits of G2 and G1 both have large eccentricities, times of pericenter that coincide to within $13$~years (a small fraction of the ${350~\mathrm{yr}}$ orbital period) and orbital angular momentum vectors that are aligned to within $20^\circ$ (i.e. a chance alignment has a probability of less than 3\%). It is therefore very likely that G1 and G2 are related objects, even though the individual orbital elements differ significantly at the level of statistical uncertainty.

\subsection{The relationship between G2 and the tail}
\label{subsec:4.3}

\begin{figure}
\centering
\includegraphics[width=0.95\linewidth]{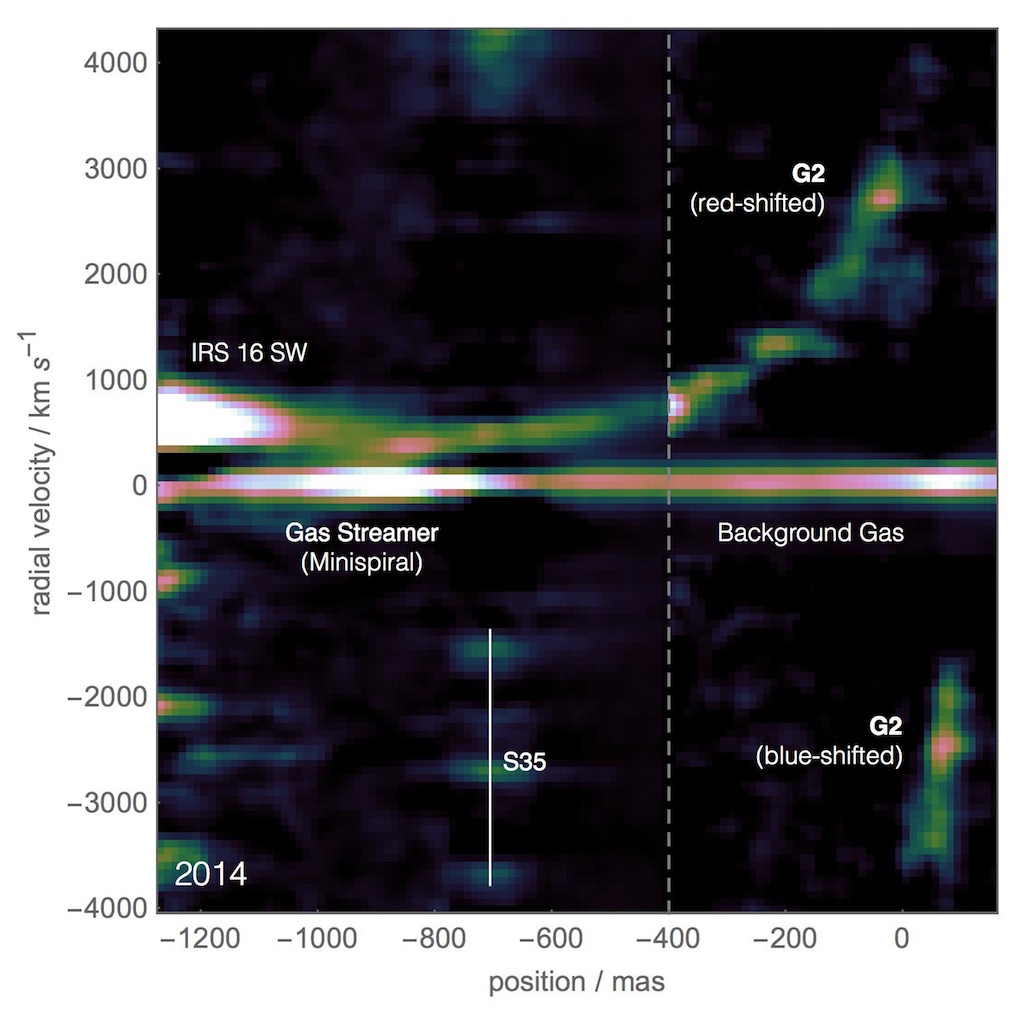}
\caption{Extended pv-diagram showing the Br$\gamma$ emission trailing G2 in 2014, which is continuously connected to the leading cloud over a large range in position and radial velocity. The color scaling on the left and right side of this graphic is adjusted separately (dashed line), to emphasize the structure of the tail rather than changes in luminosity. Where the tail intersects the star S35 in projection, the noise is increased due to the necessary subtraction of the stellar continuum emission.}
\label{fig:9}
\end{figure}

\begin{figure}
\centering
\includegraphics[width=\linewidth]{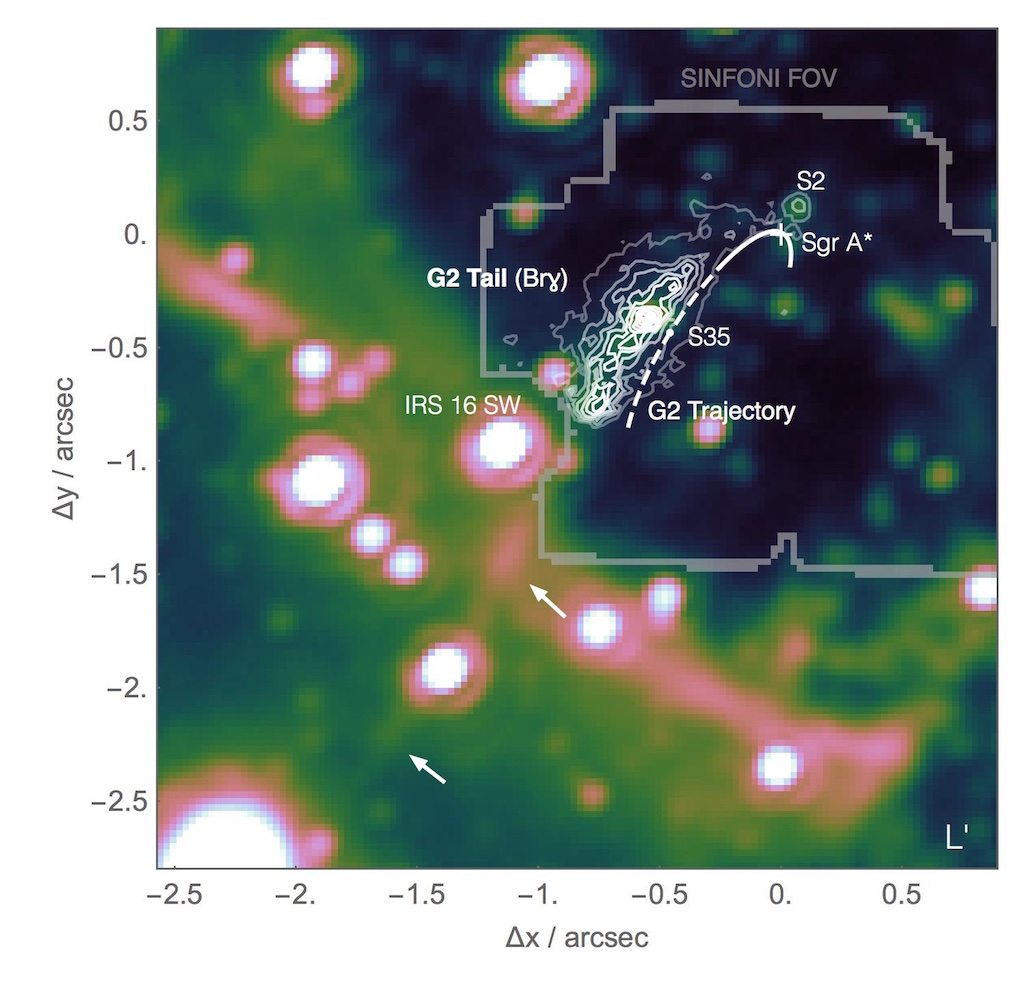}
\caption{Extended L'-band image showing the dusty filament that coincides with the tail of Br$\gamma$ emission connected to G2 (shown in contours) and which possibly continues beyond the field of view of the SINFONI observations. To visualize the full spatial extent of the Br$\gamma$ tail, which has a variable radial velocity, the data cube is collapsed along the spectral dimension using a maximum filter. Also shown is the observed trajectory of G2 (solid line) and an extrapolation of the best-fit orbit to the year 1900 (dashed line), but we note that none of the tail data has been used in this fit.}
\label{fig:10}
\end{figure}

The so-called tail of G2 is an extended filament of gas and dust trailing the main G2 source \citep[e.g.][]{2013ApJ...763...78G,2015ApJ...798..111P}, which appears to be the densest knot in this superstructure. To examine this tail in greater detail, we make use of a widely dithered set of SINFONI data that we were able to obtain as part of our 2014 observing campaign \citep{2015ApJ...798..111P}. From this data, we are able to extract an extended pv-diagram, along a trajectory that follows the apparent tail emission, which is only tentatively detected in our regular monitoring observations (Fig.~\ref{fig:9} \&~\ref{fig:10}). No particular object is known to have moved along an orbit describing such a trajectory, rather the entire tail is visible along this path at this moment in time (see also fig.~\ref{fig:12}).

The tail is continuously visible in Br$\gamma$ emission over a large range of velocities and can be traced back at least into the group of stars surrounding the eclipsing binary IRS~16~SW \citep{2006ApJ...649L.103M}, located about $1''$ towards the south-east of Sgr~A*, from where the tail appears to channel gas towards the center. There is also a dusty filament visible in L'-band that coincides with the Br$\gamma$ tail and which possibly continues even beyond the field of view of the SINFONI observations (Fig.~\ref{fig:10}). By comparison to estimates for G2, the whole tail must have a gas mass of at least several Earth masses.

The massive contact binary IRS~16~SW is a member of the clockwise stellar disk, has a broad Br$\gamma$ emission feature at a wavelength of ${2.17~\mu\mathrm{m}}$ and as a colliding-wind system could be an effective site of clump formation \citep{2016MNRAS.455.4388C} and thus a candidate origin of G2. However, the part of the tail at the very edge of the field of view does not point directly towards IRS~16~SW and it is still unclear how much further the tail might continue and whether or not it blends with the background gas.

At its other end, the tail smoothly and seamlessly connects to the red-shifted part of G2 in position-velocity space, as well as in the plane of the sky (Fig.~\ref{fig:7} \&~\ref{fig:10}), and even its velocity gradient matches that of G2 (Fig.~\ref{fig:9}), which in combination is strongly suggestive of a common origin. The tail does not lie exactly on the extrapolated orbit of G2, but a few tenths of arcseconds away in projection. At these larger distances from Sgr~A* the dynamics of the tail could well be different from those inferred for G2, as it would be expected for a genuine tidal tail \citep[e.g.][]{2014ApJ...786L..12G}. We conclude that it is very likely that both G2 and G1 are not just related to each other but to the tail as well, since again a chance occurrence would be a rare coincidence.

\section{Summary}
\label{sec:5}

During 2015 and 2016 we observed G2 completing the pericenter passage of its orbit around the supermassive black hole Sgr~A* in the Galactic Center. By putting these new data sets in the context of our previous observations, we can confidently establish the following observational facts:
\begin{itemize}
\item \textit{G2 is firmly detected as a dusty gas cloud in Br$\gamma$ line emission and L' continuum emission {\normalfont (Fig.~\ref{fig:1}, \ref{fig:2}, \ref{fig:3} \&~\ref{fig:7})}.} The spectroscopic data set, obtained with SINFONI at the ESO~VLT, spans 12 years and covers the entire pericenter passage of G2.
\item \textit{The gas cloud moves on a highly eccentric, ballistic orbit around Sgr~A* and tidally evolves approximately like a cloud of test particles between 2004 and 2016.} The gas dynamics are dominated by the gravitational force due to the black hole. The non-detection of a drag force or any strong hydrodynamic interaction with the hot gas in the inner accretion zone limits the ambient density to less than a few ${10^3~\mathrm{cm^{-3}}}$ at the pericenter distance (${1500~R_s}$), assuming G2 is a spherical cloud moving through a stationary and homogeneous atmosphere (Fig.~\ref{fig:8}). This limit is in good agreement with the non-detection of a bow shock at radio or X-ray wavelengths.
\item \textit{The gas cloud shows a pronounced spatially resolved velocity gradient both before and after pericenter {\normalfont (Fig.~\ref{fig:1} \&~\ref{fig:4})}.} This implies that the gas must be spatially extended throughout the pericenter passage.
\item \textit{The dust emission stays unresolved but co-spatial with the gas emission {\normalfont (Fig.~\ref{fig:7})}.} We note that this is consistent with the earlier observed compactness of the dust emission at around the time of pericenter \citep{2014ApJ...796L...8W}.
\item \textit{The Br$\gamma$ and L' luminosities remain constant to within the measurement uncertainty {\normalfont (Fig.~\ref{fig:5})}.} There is no need to replenish gas or dust from a potential central source.
\item \textit{G2 is likely related to G1 and the tail structure.} Although the orbital evolution of G1 and G2 is not identical, the probability of a chance occurrence is small~(Sec.~\ref{subsec:4.2}). The same is true for the tail, which is continuously connected to G2 in position-velocity space~(Sec.~\ref{subsec:4.3}).
\end{itemize}
\newpage
To further study how G2 could be used as a more powerful probe of the inner accretion zone around Sgr~A*, it will be necessary to consider more complete models of the possible interactions between G2 and the ambient medium. If they are computationally feasible, more comprehensive (hydrodynamic) simulations based on such models might be able to explain the detailed features of the observational data, which is rich in information. Moreover, it will be worth continuing to monitor the further evolution of G2, since the gas cloud might eventually interact with the ambient medium and disrupt or break up, as some hydrodynamic simulations have predicted \citep[e.g.][]{2015ApJ...811..155S}. At this point a test particle model would break down, but also give up more information about the properties of G2, as well as the accretion flow.

The continuing post-pericenter evolution of G2 over the next few years might also hold decisive clues about its nature and origin. There has been no direct detection of a stellar photosphere (in K- and H-band), but dynamic evidence for the existence of a massive central source could arise if for instance a compact dusty source would maintain Keplerian motion while the bulk of disrupting gas does not. However, in consequence G2 might also fade as a Br$\gamma$ source, resembling G1 in this respect, and become unobservable in the near future.

\acknowledgments
\noindent \textbf{Acknowledgments}\\ For helpful discussion we are grateful to A.~Ballone and M.~Schartmann, as well as J. Cuadra and D. Calder\'on.

\newpage
\appendix

\section{Supplementary Figures}

\begin{figure*}[!h]
\centering
\includegraphics[width=0.9\linewidth]{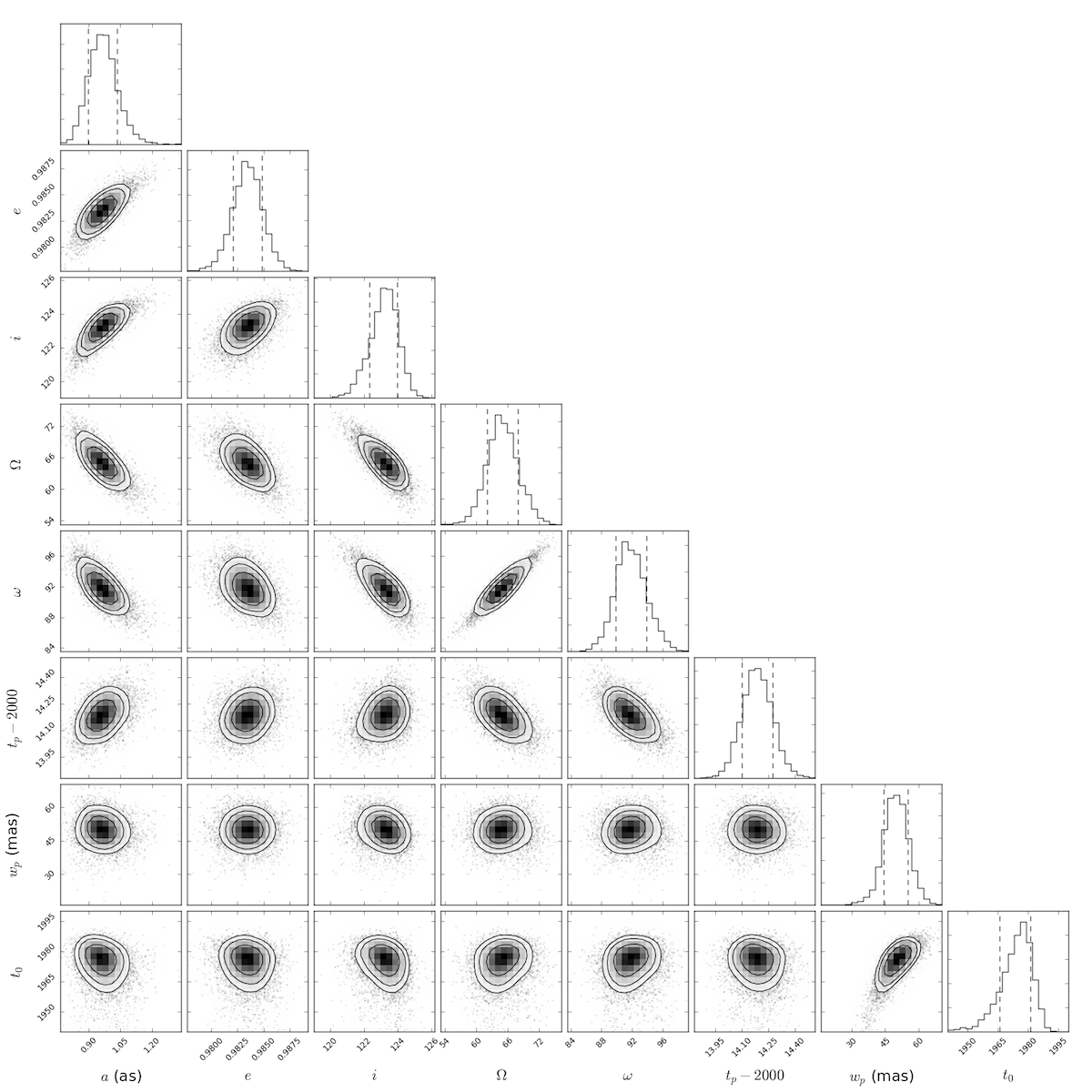}
\caption{Marginal posterior parameter distributions for the Keplerian test-particle model of G2. The dashed lines in the one-dimensional histograms indicate the 16\% and 84\% percentiles, which correspond to the innermost contour in the paired histograms.}
\label{fig:11}
\end{figure*}

\begin{figure*}[!h]
\centering
\includegraphics[width=0.4\linewidth]{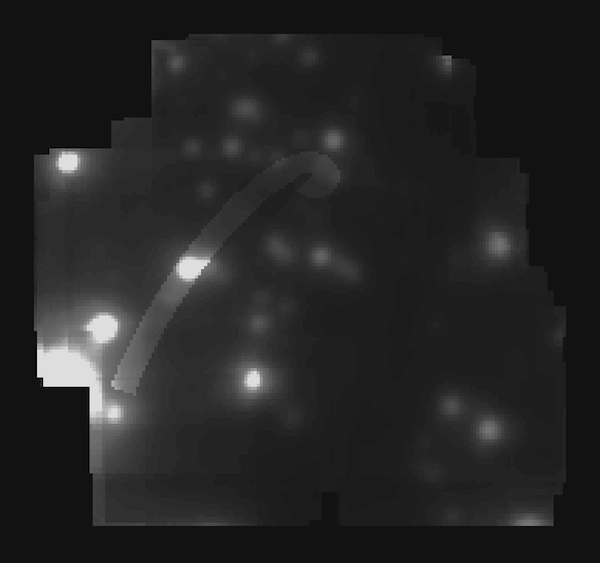}
\caption{The position of the curved, virtual slit used to extract the extended pv-diagram (see fig.~\ref{fig:9} \&~\ref{fig:10}).}
\label{fig:12}
\end{figure*}

\newpage
\bibliographystyle{abbrvnat}
\bibliography{literature}

\end{document}